# Data-Driven Approach for Uncertainty Propagation and Reachability Analysis in Dynamical Systems

Amarsagar Reddy Ramapuram Matavalam, Umesh Vaidya and Venkataramana Ajjarapu

*Abstract*—In this paper, we propose a data-driven approach for uncertainty propagation and reachability analysis in a dynamical system. The proposed approach relies on the linear lifting of a nonlinear system using linear Perron-Frobenius (P-F) and Koopman operators. The uncertainty can be characterized in terms of the moments of a probability density function. We demonstrate how the P-F and Koopman operators are used for propagating the moments. Time-series data is used for the finite-dimensional approximation of the linear operators, thereby enabling data-driven approach for moment propagation. Simulation results are presented to demonstrate the effectiveness of the proposed method.

## I. Introduction

The problem of uncertainty propagation and quantification is of interest across the various discipline of science and engineering. Examples include power system, fluid dynamics, robotics, and biological systems. In a power system, it is of interest to propagate uncertainty from renewable energy sources, various form of parametric uncertainty as well as uncertainty from the unknown initial state [1]–[3]. Characterizing set of all state that can be reached while incorporating the dynamics of the robot is essential to determine the feasibility of a path planning problem in the robotic application [4]–[6]. The problem of uncertainty propagation and reachability analysis is complicated due to the nonlinear nature of dynamics involved in these applications. Given the significance of this problem, several approaches are developed to address the uncertainty quantification problem. Among the most popular methods include Monte-Carlo based simulations, polynomial chaos, and Sum-of-Square based optimization. Each of these approaches offers some advantages and disadvantages.

A. R. R. Matavalam and V. Ajjarapu are with the Department of Electrical and Computer Engineering, Iowa State University, Ames, IA, and U. Vaidya is with the Department of Mechanical Engineering, Clemosn University, Clemson SC. uvaidya@clemson.edu

In this paper, we present some initial results on a novel approach for uncertainty propagation and reachability analysis in a dynamical system. The proposed approach relies on the linear lifting of a nonlinear system provided by linear Perron-Frobenius (P-F) and Koopman operators. While the P-F operator propagates uncertainty or probability density function under the system dynamics, the Koopman operator propagates observables [7]. These operators, therefore, provide a natural framework for the representation and propagation of uncertainty. An alternate approach of characterizing uncertainty is in terms of the moments of the probability density function. We demonstrate how the P-F and Koopman operators can be used for the propagation of moments.

By exploiting duality between P-F and Koopman operator and recent advances made in the data-driven approximation of Koopman operator, we provide an approach for the finite-dimensional approximation of P-F operator [8]–[10]. The finite-dimensional approximation of these operators is used to approximate the moments and their propagation. In particular, we introduce finite-dimensional approximation of moment propagation operator. The main contributions of the paper are as follows. The application of linear operator theoretic framework is demonstrated for the propagation of moments and the characterization of the uncertainty. A data-driven approach is proposed for the approximation and propagation of the moments. Application of the developed framework is demonstrated for moments propagation and reachability analysis on three different examples, including biological system and robotics.

## II. Moment Propagation in Nonlinear Systems Using Linear Operator

In this section we describe different but equivalent ways of lifting a nonlinear system to infinite dimensional linear system using linear operators. Consider a discrete-time deterministic dynamical system.

$$x_{t+1} = F(x_t) \tag{1}$$

where $x_t \in X$ and $F : X \to X$ is assumed to be at least $C^1$ function (i.e., differentiable function with continuous derivative). The results developed in this paper for uncertainty propagation and reachability analysis through moments can be extended to systems with random perturbations. However we restrict our analysis to deterministic system in this paper. The first approach for lifting the nonlinear dynamics is using Perron-Frobenius (P-F) operator. The P-F operator propagates uncertainty in initial conditions captured through measure or probability density function. The P-F operator on the space of measure is defined as follows.

*Definition 1 (P-F operator):* Let $\mathcal{M}(X)$ be the space of measures. The P-F operator on the space of measure $\mathbb{P}_F : \mathcal{M}(X) \to \mathcal{M}(X)$ is given by

$$[\mathbb{P}_F \mu](A) = \mu(F^{-1}(A)) \tag{2}$$

where $B \in \mathcal{B}(X)$, the Borel-$\sigma$ algebra of $X$ and $F^{-1}(B) = \{x : F(x) \in B\}$.

Under the assumption that the mapping $F$ is invertible, the P-F operator $\mathbb{P}_F : L_1(X) \to L_1(X)$ is defined as follows.

$$[\mathbb{P}_F g](x) = g(F^{-1}(x)) \left| \frac{\partial F^{-1}}{\partial x} \right|, \tag{3}$$

where $|\cdot|$ stands for the matrix determinant [1]. The second approach for lifting a nonlinear system is using Koopman operator. The Koopman operator propagates observables or functions and is defined as follows.

*Definition 2 (Koopman Operator):* Let $\mathcal{C}^0(X)$ be the space of continuous functions. The Koopman operator $\mathbb{U}_F : \mathcal{C}^0(X) \to \mathcal{C}^0(X)$ is defined as

$$[\mathbb{U}_F h](x) = h(F(x)). \tag{4}$$

The P-F and Koopman operators are dual to each other in the sense that

$$\int_X h(x) d[\mathbb{P}_F \mu](x) = \int_X [\mathbb{U}_F h](x) d\mu(x). \tag{5}$$

The duality can be expressed compactly as

$$\langle \mathbb{U}_F h, \mu \rangle = \langle h, \mathbb{P}_F \mu \rangle.$$

There is a third approach for lifting the finite dimensional nonlinear system to infinite dimensional linear system using the moment approach. Let $\mu_0(x)$ be the measure corresponding to the initial density function and $[\rho_1(x), \ldots, \rho_K(x), \ldots]$ be a choice of basis functions with respect to which the moments are computed. Moments of the initial measure $\mu_0(x)$ corresponding to these basis functions are defined as

$$m_0^k = \int_X \rho_k(x) d\mu_0(x) = \langle \rho_k, \mu_0 \rangle_X, \quad k = 1, \ldots, K, \ldots$$

The moments are propagated in time as follows.

$$m_t^k = \langle \rho_k, \mu_t \rangle = \langle \rho_k, \mathbb{P}^t \mu_0 \rangle = \langle \mathbb{U}^t \rho_k, \mu_0 \rangle$$

$$m_{t+1}^k = \langle \rho_k, \mu_{t+1} \rangle = \langle \rho_k, \mathbb{P}_F \mu_t \rangle = \langle \mathbb{U}_F \rho_k, \mu_t \rangle$$

Using the linearity of the Koopman operator the moment propagation can be expressed as

$$\mathbf{m}_{t+1} = \mathcal{K} \mathbf{m}_t$$

where $\mathbf{m}_t = [m_t^1, \ldots, m_t^K, \ldots]$ and $\mathcal{K} : \mathbb{R}^\infty \to \mathbb{R}^\infty$ is the linear operator propagating moments. In the following section we outline the procedure for the finite dimensional approximation of Koopman, P-F, and moment operators from the time-series data.

*A. Data-Driven Approximation of Linear Operators*

For the finite dimensional approximation of linear operators from data we consider projection of the infinite dimensional operators on the finitely many basis functions. We construct the finite dimensional approximation of Koopman operator from data and use it for the propagation of moments. We outline in brief the Extended Dynamic Mode Decomposition (EDMD) algorithm for the approximation of the Koopman and P-F operator. Let the time series data generated by the system (1) is

$$\mathcal{X} = [x_1, \ldots, x_N] \quad \mathcal{Y} = [y_1, \ldots, y_N]$$

where $y_k = F(x_k)$. Let the basis function used for projection is given by

$$\Psi(x) = [\varphi_1(x), \ldots, \varphi_K(x)]^\top$$

Consider any functions $\phi$ and $\hat{\phi}$ expressed as linear combination of $\Psi$ i.e.,

$$\phi(x) = \sum_{k=1}^{K} a_k \varphi_k(x) = \Psi(x)^\top \mathbf{a}$$

$$\hat{\phi}(x) = \sum_{k=1}^{K} \hat{a}_k \varphi_k(x) = \Psi(x)^\top \hat{\mathbf{a}}$$

---
[1] With some abuse of notation we are using the same notation for the P-F operator on the space of measures and functions

for some vector $\mathbf{a}, \hat{\mathbf{a}} \in \mathbb{R}^K$. The functions $\phi$ and $\hat{\phi}$ are related through Koopman as

$$\hat{\phi}(x) = [\mathbb{U}_F \phi](x) + r \qquad (6)$$

where $r$ is a residual term and arise due to the fact that the action of the Koopman is not closed on the basis functions $\Psi(x)$. Substituting for $\hat{\phi}$ and $\phi$ in Eq. (6), we obtain

$$\Psi(x)^\top \hat{\mathbf{a}} = \Psi(y)^\top \mathbf{a} + r$$

The objective is to find a matrix $\mathbf{K} \in \mathbb{R}^{K \times K}$, the finite dimensional approximation of the Koopman such that $\mathbf{Ka} = \hat{\mathbf{a}}$ such that residual $r$ is minimized. The problem of determining $\mathbf{K}$ can be written as following least square problem

$$\min_{\mathbf{K}} \| \mathbf{GK} - \mathbf{A} \|_F \qquad (7)$$

where, $\| \cdot \|_F$ stands for the Frobenius norm and

$$\mathbf{G} = \frac{1}{N} \sum_{k=1}^{N} \Psi(x_k) \Psi(y_k)^\top, \quad \mathbf{A} = \frac{1}{N} \sum_{k=1}^{N} \Psi(x_k) \Psi(x_k)^\top$$

The above minimization problem admits an analytical solution given by

$$\mathbf{K} = \mathbf{G}^\dagger \mathbf{A} \qquad (8)$$

Using the definition of the P-F operator on the space of functions or densities (3), and the duality relation between the Koopman and P-F operator can be used for the finite dimensional representation of the P-F operator. Following (5) and under the assumption that $g$ and $h$ lie in the span of $\Psi$ i.e., $g = \Psi^\top \mathbf{a}$ and $h = \Psi^\top \mathbf{b}$ for some constant vectors $\mathbf{a}$ and $\mathbf{b}$, we can write

$$\langle h, g \rangle = \mathbf{a}^\top \Lambda \mathbf{b},$$

where $[\Lambda]_{ij} = \langle \varphi_i, \varphi_j \rangle$ for $i = 2, \ldots, N$ is a symmetric matrix. We have

$$\langle \mathbb{U}_F h, g \rangle \approx (\mathbf{Kb})^\top \Lambda \mathbf{a} = \mathbf{b}^\top \mathbf{K}^\top \Lambda \mathbf{a} \approx \langle h, \mathbb{P}_F g \rangle. \qquad (9)$$

Let $\mathbf{P}$ be the finite dimensional representation of the P-F operator on the basis function, $\Psi$. Then using (9), we obtain

$$\mathbf{b}^\top \mathbf{K}^\top \Lambda \mathbf{a} \approx \langle h, \mathbb{P}_F g \rangle \approx \mathbf{b}^\top \Lambda \mathbf{P} \mathbf{a}.$$

Since the above is true for all $g$ and $h$ in the span of $\Psi(x)$, we obtain following finite dimensional approximation of the P-F operator in terms of the Koopman operator

$$\mathbf{P} = \Lambda^{-1} \mathbf{K}^\top \Lambda. \qquad (10)$$

The finite dimensional approximation of the Koopman and hence the P-F operator can be constructed from time series data and can be used for the propagation of functions under the action of P-F operator

$$\mathbf{w}_{t+1} = \mathbf{P} \mathbf{w}_t,$$

$$g_t(x) = \Psi(x)^\top \mathbf{w_t} \xrightarrow{\mathbb{P}_F} \Psi(x)^\top \mathbf{w}_{t+1} = g_{t+1}(x) \qquad (11)$$

and Koopman operator

$$\mathbf{v}_{t+1} = \mathbf{K} \mathbf{v}_t,$$

$$h_t(x) = \Psi(x)^\top \mathbf{v_t} \xrightarrow{\mathbb{U}_F} \Psi(x)^\top \mathbf{v}_{t+1} = h_{t+1}(x) \qquad (12)$$

The above two approximation of the P-F and Koopman operator can be used to study the finite dimensional propagation of moments as follows:

$$m_t^k = \langle \varphi_k, g_t \rangle \xrightarrow{\mathcal{K}} \langle \varphi_k, g_{t+1} \rangle = m_{t+1}^k \qquad (13)$$

$$m_{t+1}^k = \langle \varphi_k, \mathbb{P}_F g_t \rangle = \langle \mathbb{U}_F \varphi_k, g_t \rangle \approx \left\langle \Psi^\top \mathbf{K} e_k, g_t \right\rangle \quad (14)$$

$$= \sum_j \mathbf{K}_{jk} \langle \varphi_j, g_t \rangle = \sum_{j=1}^{K} \mathbf{K}_{jk} m_t^j \qquad (15)$$

where, $e_k$ is a column vector with all entries zero except for the $k^{th}$ entry equal to one. In the above derivation we have used the fact that $\mathbf{K}$ matrix is obtained with $\Psi(x)$ as basis functions and hence $\varphi_k(x) = \Psi(x)^\top e_k$ It then follows that

$$\mathbf{m}_{t+1} = \mathbf{K}^\top \mathbf{m}_t \qquad (16)$$

where $\mathbf{m}_t = [m_t^1, \ldots, m_t^K]^\top$ is the moment at time $t$.

## III. UNCERTAINY PROPAGATION AND REACHABILITY ANALYSIS USING MOMENT

The evolution of moments using the finite dimensional approximation of linear operator can be used for the purpose of uncertainty propagation and reachability analysis. The moment approach can be used for uncertainty propagation with uncertainty coming from initial condition as well as parametric uncertainty. In particular, for parametric uncertainty propagation, we can consider extended dynamical system with parameter as an additional state as follows.

$$\begin{aligned} z_{t+1} &= T(z_t, p_t) \\ p_{t+1} &= p_t \end{aligned} \qquad (17)$$

where $z_t$ is the true state and $p_t$ is the parameter assumed to be uncertain. Defining $x_t := (z_t^\top, p_t^\top)^\top$

and $F(x_t) := (T(x_t)^\top, p_t^\top)^\top$, we obtain dynamical system in the form (1). Let $f_0(x)$ be the probability density function capturing the uncertainty in the initial conditions and let $\Phi(x) = [\varphi_1, \ldots, \varphi_K]$ be the basis function w.r.t which the moments are computed. Then the finite moments

$$m^k = \int_X f_0(x)\varphi_k dx, \quad k = 1, \ldots, K$$

are propagated using Eq. (16). The finite dimensional approximation of the Koopman operator for moment propagation is computed with $[\varphi_1, \ldots, \varphi_K]$ as the basis functions.

For reachability analysis, one is typically interested in characterizing the set of all states that can be reached starting from given set of initial condition. The set of initial conditions can be described in multiple different ways. From example indicator functions with support on the set of initial conditions in the state space will form a natural choice for describing the reachable set. This problem can again be viewed as uncertainty propagation problem. However since we are interested in determining the reachable set, moments computed w.r.t. locally compact basis functions such as Gaussian radial basis function or indicator function will serve as a appropriate choice for moment propagation.

Let $\mathbf{m}_t$ be the moment computed with respect to Gaussian radial basis function, $\Psi(x)$, at time $t$. The reachable set, $\mathrm{RS}_t$, at time $t$ can be characterized in terms of the support of the following function i.e.,

$$\mathrm{RS}_t = \mathrm{supp}(\Psi(x)^\top \Lambda^{-1} m_t).$$

This can be explained alternatively as follows. The P-F operator is used to propagate uncertainty in initial conditions. Let $g_t(x)$ expressed as linear combinations of basis function i.e., $g_t(x) = \Psi(x)^\top w_t$ and characterizing the uncertainty in the system state at time $t$. This uncertainty is propagated forward in time using the finite dimensional approximation of the P-F operator as follows:

$$w_{t+1} = \mathbf{P} w_t. \tag{18}$$

The uncertainty in system state at time $t+1$ can then characterized using $g_{t+1}(x)$ expressed in terms of $w_{t+1}$ as follows:

$$g_{t+1}(x) = \Psi(x)^\top w_{t+1}.$$

If the uncertainty in initial condition is characterized using support of function $g_0(x)$ then the set of states that are reachable at time $t+1$ is given by $\mathrm{supp}(g_{t+1}(x))$.

Following relation can be derived between the moment $m_t \in \mathbb{R}^K$ and the vector of coefficient $w_t \in \mathbb{R}^K$,

$$w_{t+1} = \mathbf{P} w_t = \Lambda^{-1} \mathbf{K}^\top \Lambda w_t \implies \Lambda w_{t+1} = \mathbf{K}^\top \Lambda w_t.$$

Hence, the two vectors are related as follows:

$$m_t = \Lambda w_t \tag{19}$$

## IV. SIMULATION RESULTS

To validate the methodology presented in the previous sections, simulations on three dynamical systems are performed. For each of these systems, the Koopman matrix is estimated and then Monte-Carlo simulations are performed for moment propagation. The steps necessary for identifying the Koopman matrix are as follows:

1) Define a domain of interest in the state space where the state trajectories are likely to lie in.
2) Randomly select initial conditions in the domain of interest and record trajectories of the state evolution with a pre-specified time step and for a specific time duration.
3) Lift the system by calculating the value of the observables $\Psi(x)$ for each point on the recorded trajectories.
4) Estimate the Koopman matrix using the procedure outlined in Section II-A.

The steps for propagating the moments via Monte-Carlo simulation and the proposed data-driven method using the Koopman matrix are given below:

1) Sample sufficient initial conditions from a given uncertainty set and record the trajectories.
2) Estimate the probability density of the system at each time step based on the samples from the recorded trajectories.
3) Utilize the estimated probability densities to determine the moments at each time step with respect to the dictionary functions. The moment of a PDF with respect to a basis function can be calculated by numerical integration techniques. Note that the calculation of moments for the initial uncertainty set does not need any samples as the PDF is analytically known.
4) Use the equation (16) to estimate the moments at each instant progressively from the Koopman matrix. The estimated and the true moments are compared for demonstrating the accuracy.
5) In case of reachability analysis, the moments are used to reconstruct the PDFs and estimate their support as explained in Section III. This is then compared with the PDFs and their support from Monte-Carlo simulation.

`ode45` in MATLAB is used to simulate the systems and perform the Monte-Carlo simulation. The function `mvksdensity` is used to estimate the PDF from samples at various time instants. The advantage of the proposed method is that once the Koopman matrix is estimated, it can be used for various initial uncertainty sets without the need for more dynamic simulations. Further, the Monte-Carlo simulations typically need much more data than the Koopman estimation as the PDF estimation is data intensive. These make the proposed method 2-3 orders of magnitude faster for moment propagation.

### A. Example 1: Simple 2-D system

A simple 2-D system (20,21) is used to provide intuitive understanding of the method.

$$\dot{x}_1 = x_2 \tag{20}$$
$$\dot{x}_2 = -\frac{3}{2}x_1 - x_2 + \frac{x_2^3}{9} \tag{21}$$

The system has a stable equilibrium at (0,0) and the region of interest for this system is a circle around the origin with a radius of 3. 300 points are randomly selected in this region as initial points and the system is simulated from these points with a time step of 0.2s for 10s. Based on our experiments, using monomials with degree up to 2 gave good results that minimizes the error in the estimation of Koopman matrix (7). The state values are normalized by a scaling factor to ensure that the higher degree monomials do not cause numerical issues during Koopman matrix estimation. As the maximum value of each state is 3, a scaling factor of 3 is used and so the observables are:

$$\Psi(x) = \left[1, \frac{x_1}{3}, \frac{x_2}{3}, \frac{x_1^2}{9}, \frac{x_1 x_2}{9}, \frac{x_2^2}{9}\right]^\top$$

In order to test the accuracy of moment propagation, an initial uncertainty set of $(x_1, x_2) \in (-1.5, -1.1) \times (0.4, 0.8)$ is used. 1000 initial points are randomly chosen in this set and the trajectories are used to calculate the moments at each time instant. Phase portrait plot for a few sample points in the specified uncertainty set is shown in Figure 1. The moments can be estimated using the Koopman matrix and the initial uncertainty set without using any samples. Figure 2 shows the comparison between the estimated and the true moments.

It can be observed that the estimated moments closely match the true moments with a maximum error of 0.02. Furthermore, as the estimated moments did not use any samples and only uses matrix multiplication for propagating the moments, it is extremely fast. For this particular example, the Monte-Carlo simulation and PDF estimation took 2000 ms-2250 ms while the data-driven moment propagation consistently took less than 30 ms with different initial uncertainty sets. This is a speed up of 2-3 orders of magnitude.

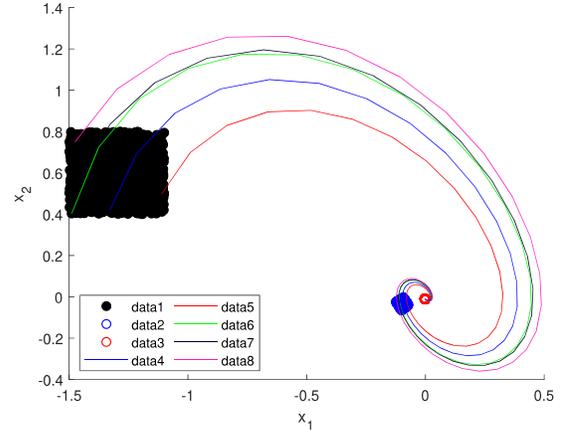

Fig. 1. Phase portrait plot for sample points and reachable sets at t=5s & t=10s for the initial uncertainty set $(-1.5, -1.1) \times (0.4, 0.8)$ in Example 1.

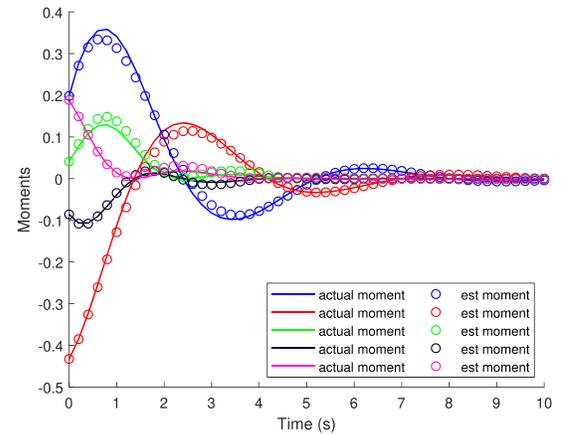

Fig. 2. The true moments and estimated moments versus time for the initial uncertainty set $(-1.5, -1.1) \times (0.4, 0.8)$ in Example 1.

### B. Example 2: Bi-stable Toggle

Next, we consider a bi-stable toggle switch system as proposed in [11]. This system models the kinetics of the concentration of two proteins which inhibit each other, resulting in two equilibrium points with complementary regions of attraction. The simplest model for this interaction is a two state repression model [11] shown in (22) where the concentration of the two proteins are denoted by $x_1$ and $x_2$.

$$\dot{x}_1 = \frac{1}{1+x_2^{3.55}} - 0.5x_1$$
$$\dot{x}_2 = \frac{1}{1+x_1^{3.53}} - 0.5x_2 \quad (22)$$

This particular example has 2 equilibrium points - (0.16,2) & (0.161,0.2). The region of interest for this system is $(x_1, x_2) \in (0, 2.5) \times (0, 2.5)$. 300 points are randomly selected in this region as initial points and the system is simulated from these points with a time step of 0.2s for 10s. Based on our experiments, using monomials with degree up to 4 (a total of 15 functions) and a scaling factor of 3 gave good results.

As there are 2 equilibrium points, two different initial uncertainty sets with different shapes are used to verify the data-driven methodology. The two initial uncertainty sets are:
- A circle centred at $(0.4, 0.8)$ with radius 0.2, shown in Figure 3. This set is in the region of attraction of $(0.16, 2)$
- A square given by $(1.2, 1.4) \times (0.5, 0.7)$ shown in Figure 5. This set is in the region of attraction of $(2, 0.161)$

The true and estimated moments for both the scenarios are plotted in Figure 4 and Figure 6. The same Koopman matrix is used for estimating the moments, even though they are in different basins of attraction. Comparing the two plots, it can be observed that different moments dominate the plots as the time advances. In case of Figure 4, the moments that are expressions of only $x_2$ rise with time (such as $x_2, x_2^2, \dots$) while the moments containing $x_1$ decay with time. This implies that the $x_2$ value increases for the samples while $x_1$ value decreases and provides an indication of how the system evolves. The reverse is true in Figure 6. It can be observed that for both the scenarios, the estimated moments closely match the true moments over the entire time with a maximum error of 0.05. Furthermore, for this particular example, the Monte-Carlo simulation (with 1000 samples) and the PDF estimation took around 7500 ms-7900 ms while the data-driven moment propagation consistently took less than 50 ms over with different initial uncertainty sets. This is again a speed up of 2-3 orders of magnitude.

## C. Example 3: Dubin Car

The Dubin car model is a classic example of non-holonomic system widely used in robotic motion planning literature. For the high level motion planning problem the dynamics of the robot are ignored. It is of interest to know if the path identified in high level planning are actually feasible for the low level planning problem where the non-holonomic dynamics of the robot are considered. This problem can be addressed by characterizing the reachable set of the robot.

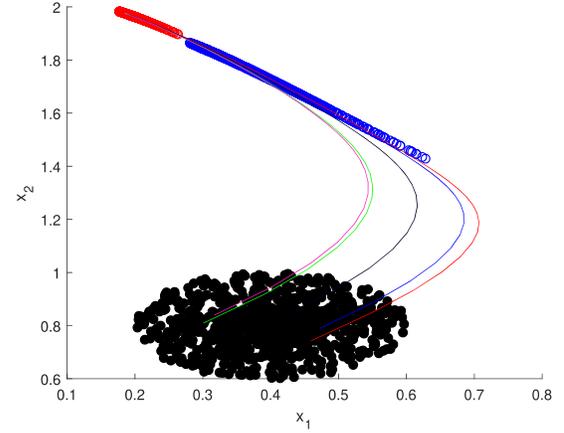

Fig. 3. Phase portrait for sample points and reachable sets at $t = 5s$ & $t = 10s$ for the initial uncertainty set in a circle centred at $(0.4, 0.8)$ with radius 0.2 in Example 2.

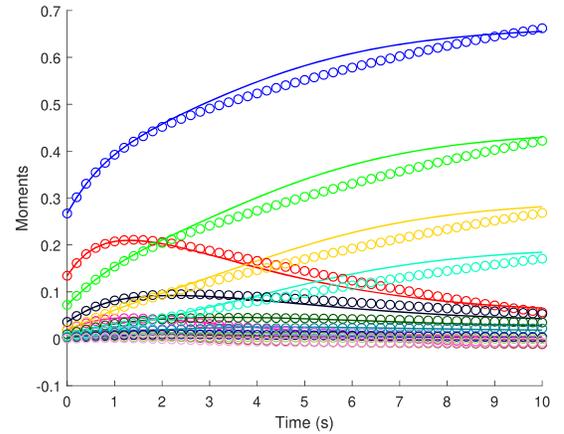

Fig. 4. The true moments and the estimated moments versus time for the initial uncertainty set in a circle centred at $(0.4, 0.8)$ with radius 0.2 in Example 2.

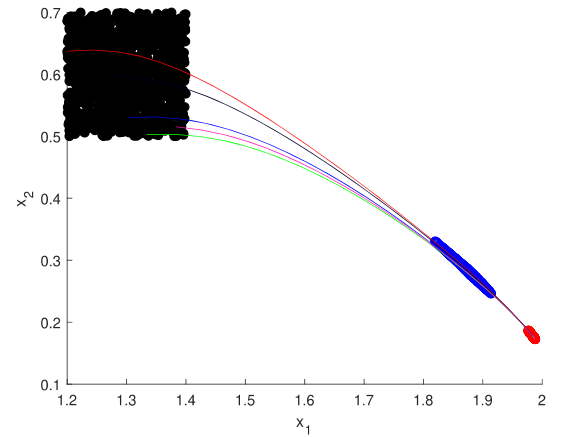

Fig. 5. Phase portrait plot for sample points and reachable sets at $t = 5s$ & $t = 10s$ for initial uncertainty set $(1.2, 1.4) \times (0.5, 0.7)$ in Example 2.

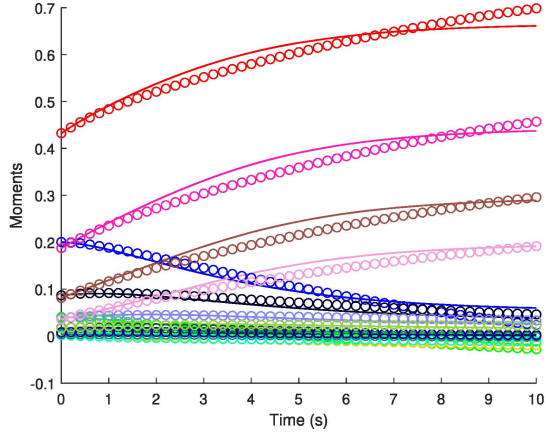

Fig. 6. The true moments and estimated moments versus time for initial uncertainty set $(1.2, 1.4) \times (0.5, 0.7)$ in Example 2.

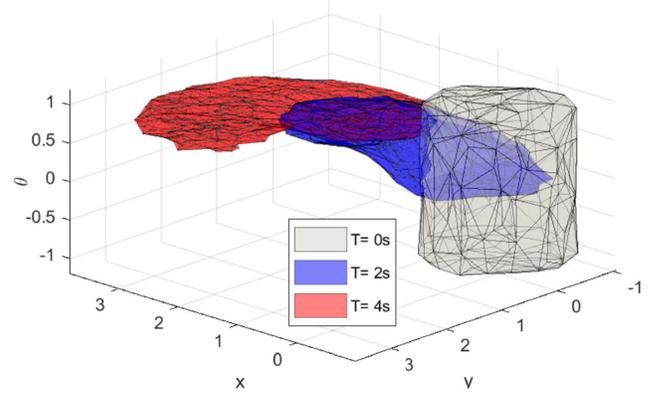

Fig. 7. The initial uncertainty set along with the reachable sets at at $t = 2s$ and $t = 4s$ for Example 3.

$$\dot{x} = \nu \cos \theta \quad (23)$$
$$\dot{y} = \nu \sin \theta \quad (24)$$
$$\dot{\theta} = \omega \quad (25)$$

$\nu$ and $\omega$ are control inputs. Consider following feedback control input.

$$\begin{bmatrix} \nu \\ \omega \end{bmatrix} = \begin{bmatrix} \nu_{dx} cos(\theta) + \nu_{dy} sin(\theta) \\ \frac{1}{b}(\nu_{dy} cos(\theta) - \nu_{dx} sin(\theta)) \end{bmatrix} \quad (26)$$

where $\nu_{dx} = \nu_{dy} = 0.6$ and $b = 2$. This system does not have an equilibrium and we are interested in performing a reachability analysis by utilizing the moments to identify the regions where the trajectories of the states will lie after starting from an initial uncertainty set.

The region of interest for this system is $(x, y, \theta) \in (-4, 6) \times (-4, 6) \times (-1.5, 1.5)$. 1000 points are randomly selected in this region as initial points and the system is simulated with a time step of 0.2s for 4s. For performing reachability analysis, 1-D Gaussian radial basis functions (RBFs) are used as dictionary functions with their centers equally spaced on the individual axis in the domain of interest. We used 12 RBFs with their centers equally spaced along each axis (a total of 36 dictionary functions). As 1-D RBFs are used, the projection of the reachable set along each axis can be estimated from the dictionary functions.

In order to test the accuracy of proposed data-driven reachability estimation, an initial uncertainty set of a cylinder of height 2 centered at the origin along the $\theta$-axis with a radius of 1 is used. 3000 initial points are uniformly chosen in this set and the trajectories are used to calculate the moments at each time instant. The initial uncertainty set along with the reachable sets at $t = 2s$ and $t = 4s$ from the Monte-Carlo simulation are shown in Figure 7.

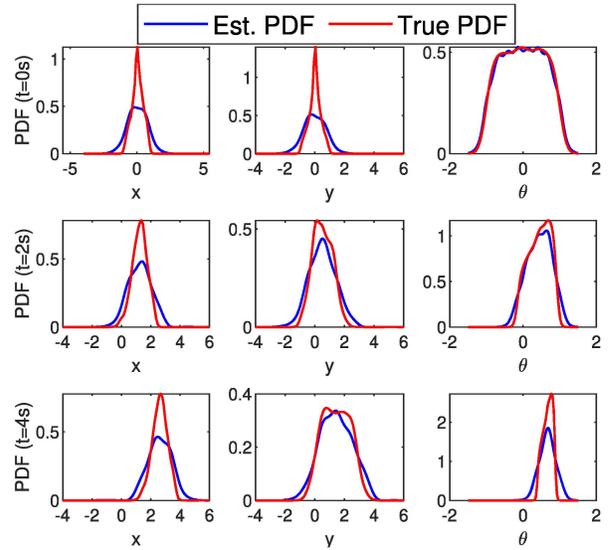

Fig. 8. The projection of the true PDF and the estimated PDF along the x, y & $\theta$ axis at $t = 0s, 2s \& 4s$ for the initial uncertainty set in Example 3.

It can be seen from Figure 7 that the range of $\theta$ state decreases with time and tends to converge to a fixed value. Thus, the height of the reachable set decreases as the system evolves. Thus, the probability density function (PDF) of the reachable set projected along the $\theta$ axis will be supported over a smaller region as the system evolves, which increases the peak of the PDF. The next observation is that the range of the x and y states becomes larger with time and so the reachable set spreads out more in the x-y plane as the system evolves. Thus, along these axis, the support of the projected PDF will be increasing over time. Furthermore, the evolution of the y-state is faster than the x-state and so the support of the projected PDF along y-axis will be larger than the support along x-axis.

Figure 8 plots the true PDF and the PDF estimated from the moments as explained in Section III for $t = 0s, 2s$ and $4s$. It can be observed from the plots at $t = 0s$, that the support of the estimated PDF along the x and y axis are conservative with respect to the true PDF- i.e. the support of the estimated PDF is a super-set of the true support. This in not observed in the PDF plot of $\theta$ at $t = 0s$ as there is a good match between the true and estimated PDF. This mismatch is due to the fact that the dictionary functions are wider along the x and y dimensions compared to the $\theta$ dimension (around 3 times wider). The support of the initial uncertainty along the x and y directions are not large enough to be well estimated by a linear combination of RBFs that have finite support. Thus, the best approximation is a single RBF function that is scaled to make the bounded area equal to 1. In case of the PDF of $\theta$, the initial support is large enough to be spanned by several RBFs and thus it can be approximated much better.

As the system evolves, the support of x and y increases and so the estimated PDF is able to better approximate the true PDF and its error reduces with time. Furthermore, the error in the estimated support (PDF) of y reduces faster than the error in the support (PDF) of x and this is because the support of y increases faster than x, as explained previously. Conversely, as the support of $\theta$ reduces rapidly as the system evolves, it can be seen that the estimated PDF becomes more erroneous as the system evolves. However, even when the estimated PDF is erroneous, the estimated support is a super-set of the true support. Finally, for this particular example, the Monte-Carlo simulation took around 15 s-15.7 s while the data-driven moment propagation and PDF estimated consistently took less than 120 ms with different initial uncertainty sets. This is a speed up of 2-3 orders of magnitude.

## V. Conclusions and Future Direction

We presented some preliminary results on the use of linear operator theoretic framework for uncertainty propagation and reachability analysis in dynamical system from time-series data. The proposed method is data-driven and hence does not require detailed system model. Detailed comparison with other model-free methods such as Monte-Carlo on the numerical efficiency of the two approaches will be the topic of future investigation. We are in the process of extending these methods to power systems to analyze how uncertainty due to renewable resources and parametric uncertainty impacts the dynamic performance of the system after a fault.

## References

[1] Y. Xu, L. Mili, A. Sandu, M. R. von Spakovsky, and J. Zhao, "Propagating uncertainty in power system dynamic simulations using polynomial chaos," *IEEE Transactions on Power Systems*, vol. 34, no. 1, pp. 338–348, 2018.

[2] H. Choi, P. J. Seiler, and S. V. Dhople, "Propagating uncertainty in power-system dae models with semidefinite programming," *IEEE Transactions on Power Systems*, vol. 32, no. 4, pp. 3146–3156, 2016.

[3] D. A. Maldonado, M. Schanen, and M. Anitescu, "Uncertainty propagation in power system dynamics with the method of moments," pp. 1–5, Aug 2018.

[4] J. H. Gillula, G. M. Hoffmann, H. Huang, M. P. Vitus, and C. J. Tomlin, "Applications of hybrid reachability analysis to robotic aerial vehicles," *The International Journal of Robotics Research*, vol. 30, no. 3, pp. 335–354, 2011.

[5] M. Althoff and J. M. Dolan, "Online verification of automated road vehicles using reachability analysis," *IEEE Transactions on Robotics*, vol. 30, no. 4, pp. 903–918, 2014.

[6] J. Ding, E. Li, H. Huang, and C. J. Tomlin, "Reachability-based synthesis of feedback policies for motion planning under bounded disturbances," in *2011 IEEE International Conference on Robotics and Automation*. IEEE, 2011, pp. 2160–2165.

[7] A. Lasota and M. C. Mackey, *Chaos, Fractals, and Noise: Stochastic Aspects of Dynamics*. New York: Springer-Verlag, 1994.

[8] C. W. Rowley, I. Mezić, S. Bagheri, P. Schlatter, and D. S. Henningson, "Spectral analysis of nonlinear flows," *Journal of fluid mechanics*, vol. 641, pp. 115–127, 2009.

[9] M. O. Williams, I. G. Kevrekidis, and C. W. Rowley, "A data-driven approximation of the koopman operator: Extending dynamic mode decomposition," *Journal of Nonlinear Science*, vol. 25, no. 6, pp. 1307–1346, 2015.

[10] B. Huang and U. Vaidya, "Data-driven approximation of transfer operators: Naturally structured dynamic mode decomposition," in *2018 Annual American Control Conference (ACC)*. IEEE, 2018, pp. 5659–5664.

[11] T. S. Gardner, C. R. Cantor, and J. J. Collins, "Construction of a genetic toggle switch in escherichia coli," *Nature*, vol. 403, no. 1, pp. 339–342, 2000.